\documentclass{amsart}
\usepackage{graphicx}
\newtheorem{thm}{Theorem}[section]
\newtheorem{cor}[thm]{Corollary}
\newtheorem{lem}[thm]{Lemma}
\newtheorem{prop}[thm]{Proposition}
\theoremstyle{definition}
\newtheorem{defn}[thm]{Definition}
\theoremstyle{remark}

\numberwithin{equation}{section}

\begin{document}

\title[Exponential sums, Nowton identities and Dickson polynomials]{On Exponential Sums, Nowton identities and Dickson Polynomials over Finite Fields}%
\author[Xiwang Cao, Lei Hu]
{Xiwang Cao$^1$, Lei Hu}
\thanks{$^1$Research supported by NNSF Grant 10971250, 10771100.}

\address{Xiwang Cao is with School of Mathematical Sciences, Nanjing University of
Aeronautics and Astronautics, Nanjing 210016, email: {\tt
xwcao@nuaa.edu.cn}}

\address{Lei Hu is with State Key State Lab.
of Information Security, Graduate School of Chinese Academy of
Sciences, Beijing 100049, P. R. China, email: {\tt hu@is.ac.cn}}

\begin{abstract}
Let $\mathbb{F}_{q}$ be a finite field, $\mathbb{F}_{q^s}$ be an extension of $\mathbb{F}_q$, let $f(x)\in \mathbb{F}_q[x]$ be a polynomial of degree $n$ with $\gcd(n,q)=1$. We present a recursive formula for evaluating the exponential sum $\sum_{c\in \mathbb{F}_{q^s}}\chi^{(s)}(f(x))$. Let $a$ and $b$ be two elements in $\mathbb{F}_q$ with $a\neq 0$, $u$ be a positive integer. We obtain an estimate for the exponential sum $\sum_{c\in \mathbb{F}^*_{q^s}}\chi^{(s)}(ac^u+bc^{-1})$, where $\chi^{(s)}$ is the lifting of an additive character $\chi$ of $\mathbb{F}_q$. Some properties of the sequences constructed from these exponential sums are provided also.

\end{abstract}
\maketitle
\section{Introduction}

Let $\mathbb{F}_q$ denote the finite field of characteristic $p$
with $q$ elements ($q=p^e, e\in \mathbb{N}$, the set of positive integers), and $\mathbb{F}_q^*$
the non-zero elements of $\mathbb{F}_q$. Let $\mathbb{F}_q[x]$ be the polynomial ring with indeterminate $x$. For every positive integer $s$ and a positive divisor $t$ of $s$, the {\it relative trace map} from $\mathbb{F}_{q^s}$ to $\mathbb{F}_{q^t}$ is defined as
\begin{equation}\label{f-1}
   {\rm Tr}^s_t(c)=c+c^{2^t}+c^{2^{2t}}+\cdots+c^{2^{s-t}}, \forall c\in \mathbb{F}_{q^s}.
\end{equation}
The {\it absolute trace map} is defined by
\begin{equation*}
    {\rm Tr}(c)=c+c^p+c^{p^2}+\cdots+c^{p^{e-1}}, \forall c\in \mathbb{F}_q,
\end{equation*}
and the function maps from $\mathbb{F}_q$ to $\mathcal{C}^*$, the set of nonzero complex numbers, defined by
\begin{equation*}
    \chi_a(c)=e^{2\pi \sqrt{-1}{\rm Tr}(ac)/p}, \forall c\in \mathbb{F}_q,
\end{equation*}
is called an {\it additive character} of  $\mathbb{F}_q$.
Let $\chi$ be an additive character of $\mathbb{F}_q$, and $f(x)\in \mathbb{F}_q[x]$. The sum
\begin{equation*}
    \mathcal{S}(f)=\sum_{c\in \mathbb{F}_q}\chi(f(c))
\end{equation*}
is called a {\it Weil Sum}.
Let $g$ be a generator of the cyclic group $\mathbb{F}_q^*$, the function maps from $\mathbb{F}_q$ to $\mathcal{C}^*$ defined by
\begin{equation*}
   \psi(g^k)=e^{2\pi \sqrt{-1}k/(q-1)}, \mbox{ for } k=0,1,2,\cdots,q-2,
\end{equation*}
is called a {\it multiplicative character} of $\mathbb{F}_q$. It is easy to see that $\psi$ is a generator of the characteristic group of $\mathbb{F}_q^*$. For every multiplicative character $\psi$ of $\mathbb{F}_q$ and a polynomial $f(x)\in \mathbb{F}_q[x]$, one can also define the following exponential sum
\begin{equation*}
   \mathcal{T}(f)=\sum_{c\in \mathbb{F}_q}\psi(f(c)),
\end{equation*}
here we extend the definition of $\psi$ to the set $\mathbb{F}_q$ by setting $\psi(0)=0$.

The problem of
explicitly evaluating these sums, $\mathcal{S}(f), \mathcal{T}(f)$, is quite often difficult. Results giving estimates
for the absolute value of the sums are more common and such results have been
regularly appearing for many years. Lidl and Niederreiter gave an
overview of this area of research in the concluding remarks of Chapter 5 in \cite{lidl}. See also, \cite{coulter,lison, moiso,vlugt,wan,weil1,weil2} for instance.

Using the technique of L-functions, one can prove the following results:

{\thm \label{thm-1}\cite[p.~220, Theorem 5.36]{lidl} Let $f(x)\in \mathbb{F}_q[x]$ be of degree $n\geq 2$ with $\gcd(n,q)=1$ and let $\chi$ be a nontrivial additive character of $\mathbb{F}_q$. Then there exist complex numbers $\omega_1,\omega_2,\cdots, \omega_{n-1}$, only depending on $f$ and $\chi$, such that for any positive integer $s$ we have
\begin{equation}\label{f-2}
    \sum_{\gamma\in \mathbb{F}_{q^s}}\chi^{(s)}(f(\gamma))=-\omega_1^s-\omega_2^s-\cdots\omega_{n-1}^s,
\end{equation}
where $\chi^{(s)}(x)=\chi({\rm Tr}^s_1(x))$ for all $x\in \mathbb{F}_{q^s}$ is the lifting of $\chi$ from $\mathbb{F}_q$ to $\mathbb{F}_{q^s}$.}

{\thm \label{thm-2} \cite[p.223, Theorem 5.39]{lidl} Let $\psi$ be a multiplicative character of $\mathbb{F}_q$ of order $m>1$ and $f\in \mathbb{F}_q[x]$ be a monic polynomial of positive degree that is not an $m$-th power of a polynomial. Let $d$ be the number of distinct roots of $f$ in its splitting field over $\mathbb{F}_q$ and suppose that $d\geq 2$. Then there exist complex numbers $\theta_1,\theta_2,\cdots,\theta_{d-1}$, only depending on $f$ and $\psi$, such that for every positive integer $s$ we have
\begin{equation*}
   \sum_{\gamma\in \mathbb{F}_{q^s}}\psi^{(s)}(f(\gamma))=-\theta_1^s-\theta_2^s-\cdots-\theta_{d-1}^s,
\end{equation*}
where $\psi^s(x)=\psi({\rm Norm}_1^s(x))$ for all $x\in \mathbb{F}_{q^s}$ is the lifting of $\psi$ from $\mathbb{F}_q$ to $\mathbb{F}_{q^s}$, and
\begin{equation*}
    {\rm Norm}_1^s(x)=x^{1+q+q^2+\cdots+q^{s-1}}.
\end{equation*}}

For convenience, we make the following notations:
\begin{equation}\label{f-0}
    \mathcal{S}_s(f)= \sum_{\gamma\in \mathbb{F}_{q^s}}\chi^{(s)}(f(\gamma)), \mathcal{T}_s(f)= \sum_{\gamma\in \mathbb{F}_{q^s}}\psi^{(s)}(f(\gamma)).
\end{equation}

In this note, we establish some recursive formulae about $\mathcal{S}_s(f)$ and $\mathcal{T}_s(f)$. Some results abut the exponential sum $G_u^{(s)}(a,b)=\sum_{c\in \mathbb{F}^*_{q^s}}\chi^{(s)}(ac^u+bc^{-1})$ are obtained, where $u$ is a positive integer, $a,b\in \mathbb{F}_q$.

The organization of the rest of the paper is as follows: In Sect. 2, we introduce some preliminaries results which will be used in the sequel. In Sect. 3, we give a recursive formula and an estimate for the exponential sum $G_u(a,b)=\sum_{c\in \mathbb{F}_q^*}\chi(ac+bc^{-1})$. In Sect. 4, we provide some properties of the sequences constructed from these exponential sums.
\section{Preliminaries}

In this section, we introduce the concept of Dickson polynomials, Newton's identities and an useful tool for evaluating exponential sums.

\subsection{Dickson polynomials}

Before going to state our main results, we need the concept of Dickson polynomials.

Consider a polynomial

\begin{equation}\label{f-3}
    r(c_1,c_2,\cdots,c_k,x)=x^{k+1}-c_1x^k+c_2x^{k-1}+\cdots+(-1)^kc_kx+(-1)^{k+1}a\in \mathbb{F}_q[x].
\end{equation}
This polynomial has $k+1$ not necessarily distinct roots $\beta_1,\cdots,\beta_{k+1}$ in a suitable extension of $\mathbb{F}_q$. Now, let $n\in \mathbb{N}$, the set of positive integers, and set
\begin{equation*}
  r_n(c_1,\cdots,c_k,x)=(x-\beta_1^n)\cdots (x-\beta_{k+1}^n).
\end{equation*}
Define
\begin{eqnarray*}
e_0(x_1,x_2,\cdots,x_{k+1})&=&1\\
  e_1(x_1,x_2,\cdots,x_{k+1}) &=& \sum_{i=1}^{k+1}x_i \\
  e_2(x_1,x_2,\cdots,x_{k+1})  &=& \sum_{1\leq i<j\leq k+1}^{k+1}x_ix_j \\
 \dots &\cdots& \cdots \\
 e_{k+1}(x_1,x_2,\cdots,x_{k+1})  &=& x_1x_2\cdots x_{k+1}\\
 e_m(x_1,x_2,\cdots,x_{k+1})  &=&0, \mbox{ for }m>k+1.
\end{eqnarray*}
We know that the coefficients of $r_n$ are elementary symmetric functions $e_i(\beta_1^n,\cdots,\beta_{k+1}^n),$ $i=1,2,\cdots,k+1$. Since $e_i$ is symmetric in the indeterminates $x_1,\cdots,x_{k+1}$, there exist integral polynomials $D^{(1)}_n, \cdots, D_n^{(k+1)}$ in $k+1$ indeterminates such that
\begin{equation*}
   e_i(x_1^n,\cdots, x_{k+1}^n)=D_n^{(i)}(e_1(x_1,\cdots,x_{k+1}),\cdots,e_{k+1}(x_1,\cdots,x_{k+1}))
\end{equation*}
for $1\leq i\leq k+1$.
Since $\beta_1,\cdots,\beta_{k+1}$ are the roots of the polynomial $r$ we have $e_i(\beta_1,\cdots,\beta_{k+1})=c_i$ for $1\leq i\leq k$ and $e_{k+1}(\beta_1,\cdots, \beta_{k+1})=a$. Thus we have
\begin{equation*}
   D_n^{(i)}(c_1,\cdots,c_k,a)=e_i(\beta_1^n,\cdots,\beta_{k+1}^n).
\end{equation*}
Therefore,
\begin{equation*}
    r_n(c_1,\cdots,c_k,x)=x^{k+1}+\sum_{1\leq i\leq k}(-1)^iD_n^{(i)}(c_1,\cdots,c_k,a)x^{k+1-i}+(-1)^{k+1}a^n.
\end{equation*}

\begin{defn} The Dickson polynomials of the first kind $D_n^{(i)}(x_1,\cdots,x_k,a)$, $1\leq i\leq k$, are given by the functional equations
\begin{equation*}
   D_n^{(i)}(x_1,\cdots,x_k,a)=e_i(u_1^n,\cdots,u_{k+1}^n), 1\leq i\leq k,
\end{equation*}
where $x_i=e_i(u_1,\cdots,u_{k+1})$ and $u_1\cdots u_{k+1}=a$.\end{defn}

Thus, particularly, for $i=1$, we have
\begin{equation}\label{f-4}
    D_n^{(1)}(e_1,e_2,\cdots, e_{k+1})=u_1^n+u_2^n+\cdots+u_{k+1}^n,
\end{equation}
where $e_j=e_j(u_1,\cdots,u_{k+1})$.

Waring's formula also gives the explicit expression of $D_n^{(1)}(x_1,\cdots,x_k,a)$ as
\begin{eqnarray*}
    &&D_n^{(1)}(x_1,\cdots,x_k,a) \\
    &=& \sum_{i_1=0}^{\lfloor \frac{n}{2}\rfloor}\cdots \sum_{i_k=0}^{\lfloor \frac{n}{k+1}\rfloor}\frac{n}{n-i_1-2i_2-\cdots-ki_k}{n-i_1-2i_2-\cdots-ki_k\choose i_1+\cdots+i_{k-1}}\\
    &&{i_1+\cdots+i_k\choose i_1+\cdots+i_{k-1}}\cdots{i_1+i_2\choose i_1}a^{i_k}(-1)^{i_1+2i_2+\cdots+ki_k}\\
    &&x_1^{n-2i_1-\cdots-(k+1)i_k}x_2^{i_1}\cdots x_k^{i_{k-1}},
\end{eqnarray*}
where $\lfloor x \rfloor$ denotes the largest integer $\leq x$. See Lidl \cite[~p.19]{LMT} for details.

Moreover, we have the following results.

{\lem \cite[~p.19]{LMT}The Dickson polynomials of the first kind $D_n^{(1)}(x_1,\cdots,x_k,a)$ satisfy the generating function
\begin{equation*}
   \sum_{n=0}^{\infty}D_n^{(1)}(x_1,\cdots,x_k,a)z^n=\frac{\sum_{i=0}^k(k+1-i)(-1)^ix_iz^i}{\sum_{i=0}^{k+1}(-1)^ix_iz^i} \mbox{ for $n\geq 0$}
\end{equation*}
and the recurrence relation
\begin{equation}\label{f-5}
  D_{n+k+1}^{(1)}-x_1D_{n+k}^{(1)}+\cdots+(-1)^kx_kD_{n+1}^{(1)}+(-1)^{k+1}aD_n^{(1)}=0
\end{equation}
with the $k+1$ initial values
\begin{equation*}
   D_0^{(1)}=k+1,D_j^{(1)}=\sum_{t=1}^j(-1)^{t-1}x_tD_{j-t}^{(1)}+(-1)^j(k+1-j)x_j \quad\mbox{for $0<j\leq k$}.
\end{equation*}
}

\subsection{Newton's identities}
Let $x_1,\cdots, x_{k+1}$ be $k+1$ unnecessarily distinct numbers, denote for $n \geq 1$ by $p_n(x_1,\cdots,x_{k+1})$ the $n$-th power sum:
\begin{equation*}
    p_n(x_1,\cdots,x_{k+1})=x_1^n+x_2^n+\cdots+x_{k+1}^n.
\end{equation*}
Then Newton's identities can be stated as
\begin{equation}\label{f-6}
    me_m(x_1,\cdots,x_{k+1})=\sum_{j=1}^m(-1)^{j-1}e_{m-j}(x_1,\cdots,x_{k+1})p_j(x_1,\cdots,x_{k+1})
\end{equation}
valid for all $m\geq 1$. See ${\rm http://en.wikipedia.\ org/ wiki/ Newton's\ identities}$ for details.

For convenience, we denote $p_j(x_1,\cdots,x_{k+1})$ simply by $p_j$, and denote $e_j(x_1,\cdots,x_{k+1})$ by $e_j$.
By (\ref{f-6}), we know that
\begin{equation}\label{f-7}
   p_m=\sum_{j=1}^{m-1}(-1)^{j-1}e_jp_{m-j}+(-1)^{m-1}me_m, m\geq 2.
\end{equation}

It is easily seen that both (\ref{f-7}) and (\ref{f-5}) tell the something.

By (\ref{f-7}) and Cramer's rule, we obtain  that
\begin{eqnarray}\label{f-8}
 p_m=\left|\begin{array}{ccccc}
             e_1 & 1 & 0 & \cdots &  \\
            2e_2 & e_1 & 1 & 0& \cdots \\
             3e_3 & e_2 & e_1 &\ddots &  \\
             \vdots & \vdots & \vdots & \ddots &  \\
             me_m & e_{m-1} & \cdots & & e_1
           \end{array}
 \right|, \end{eqnarray}and\begin{eqnarray}\label{f-9}e_m=\frac{1}{m!}\left|\begin{array}{ccccc}
             p_1 & 1 & 0 & \cdots &  \\
            p_2 & p_1 & 2 & 0& \cdots \\
             \vdots & \vdots & \ddots& \ddots &  \\
             p_{m-1}&p_{m-2}&\cdots&p_1&n-1\\
             p_m & p_{m-1} & \cdots & p_2& p_1
           \end{array}\right|.
\end{eqnarray}

As an application of Newton's identities, we show that (\ref{f-5}) is sufficient to define the Dickson polynomial of the first kind $D_n^{(1)}$.

{\prop If $D_n$ satisfies (\ref{f-5}) with the initial values, then as a sequence, $D_n$ is just $D_n^{(1)}$.}

 \begin{proof}
 As a sequence, $D_n$ has the characteristic polynomial as
\begin{equation*}
     r(x_1,x_2,\cdots,x_k,x)=x^{k+1}-x_1x^k+x_2x^{k-1}+\cdots+(-1)^kx_kx+(-1)^{k+1}a.
\end{equation*}
Let the roots of $ r(x_1,x_2,\cdots,x_k,x)=0$ be $\beta_1,\cdots,\beta_{k+1}$. Then we have $x_i=e_i(\beta_1,\cdots,\beta_{k+1})$, and for every positive integer $n$ $$D_n^{(1)}=l_1\beta_1^n+\cdots+l_{k+1}\beta_{k+1}^n$$ holds for some constants $l_1,\cdots,l_{k+1}$. Now by the initial values of $D_j^{(1)}$, we get the following system of equations.
\begin{equation*}
\left\{ \begin{array}{ll}
 l_1+\cdots+l_{k+1} = k+1& \\
  \sum_{v=1}^{k+1}l_v\beta_v^j=\sum_{t=1}^j(-1)^{t-1}x_t\sum_{v=1}^{k+1}l_v\beta_v^{j-t}+(-1)^j(k+1-j)x_j \quad\mbox{for $0<j\leq k$}.&
 \end{array}\right.
\end{equation*}
Simplifying this system, we have
\begin{equation*}
\left\{ \begin{array}{ll}
 l_1+\cdots+l_{k+1} = k+1& \\
  jx_j=\sum_{t=1}^j(-1)^{t-1}x_{j-t}\sum_{v=1}^{k+1}l_v\beta_v^t\quad\mbox{for $0<j\leq k$}.&
 \end{array}\right.
\end{equation*}
Comparing with (\ref{f-6}), we obtain $l_1=\cdots=l_{k+1}=1$ is a solution to this system. Furthermore, $D_n$ is uniquely determined by its initial values. Thus we obtain $l_1=\cdots=l_{k+1}=1$ is the unique solution to this system and $D^{(1)}=\beta_1^n+\cdots+\beta_{k+1}^n$.
This completes the proof.\end{proof}

Thus we obtain an equivalent definition of Dickson polynomials of the first kind.

Now we consider the exponential sums defined in Theorem \ref{thm-1} and Theorem \ref{thm-2}. Using the relations between the power sums and the elementary functions, we obtain the following recursive formulae:

{\prop (1) Let $f$ and $\mathcal{S}_s(f)$ be defined as in Theorem \ref{thm-1} and (\ref{f-0}).Then for every integer $s\geq 2$, we have
\begin{equation*}
   \mathcal{S}_s(f)=\sum_{j=1}^{s-1}(-1)^{j-1}e_{j}(\omega_1,\cdots,\omega_{n-1})\mathcal{S}_{s-j}(f)+(-1)^sse_s(\omega_1,\cdots,\omega_{n-1}).
\end{equation*}
(2) Let $f$ and $\mathcal{T}_s(f)$ be defined as in Theorem \ref{thm-2} and (\ref{f-0}). Then for every integer $s\geq 2$, we have
\begin{equation*}
    \mathcal{T}_s(f)=\sum_{j=1}^{s-1}(-1)^{j-1}e_{j}(\theta_1,\cdots,\theta_{d-1})\mathcal{T}_{s-j}(f)+(-1)^sse_s(\theta_1,\cdots,\theta_{d-1}).
\end{equation*}

}

\subsection{An useful tool for computing exponential sums}
In order to determine some exponential sums over finite fields, we use the idea of $L-$functions: The canonical L-function of the exponential sum $\mathcal{S}_s(f)=\sum_{c\in \mathbb{F}_{q^s}}\chi^{(s)}(f(c))$ is defined by $L^*(f,t)=\exp\left(\sum_{s=1}^{\infty}\mathcal{S}_s(f)\frac{t^s}{s}\right)$. By a famous theorem of Dwork and Grothendieck, $L^*(f, t)$ is a rational function.

Let $\Phi$ be the set of monic polynomials over $\mathbb{F}_q$, and let $\lambda$ be a complex-valued function on $\Phi$ which is
multiplicative in the sense that
\begin{equation}\label{f-10}
    \lambda(gh)=\lambda(g)\lambda(h) \mbox{ for all }g,h\in\Phi,
\end{equation}
and which satisfies $|\lambda(g)|\leq 1$ for all $g\in \Phi$, and $\lambda(1)=1$. Denote by $\Phi_k$ the subset of $\Phi$ containing the polynomials of degree $k$.

Consider the power series
\begin{equation}\label{f-11}
    L(z)=\sum_{k=0}^{\infty}\left(\sum_{g\in \Phi_k}\lambda(g)\right)z^k.
\end{equation}
It is easily seen that
\begin{equation*}
    L(z)=\prod_{f\ irr}(1-\lambda(f)z^{\deg(f)})^{-1},
\end{equation*}
where the product is taken over all monic irreducible polynomials $f$ in $\mathbb{F}_q[x]$. Now apply logarithmic differentiation and multiply by $z$ to get
\begin{equation}\label{f-12}
    z\frac{d\log L(z)}{dz}=\sum_{s=1}^{\infty}L_sz^s
\end{equation}
with
\begin{equation}\label{f-13}
   L_s=\sum_{f}\deg(f)\lambda(f)z^{s/\deg(f)}
\end{equation}
where the sum is extended over all monic irreducible polynomials $f$ in $\mathbb{F}_q[x]$ with degree dividing $s$.

Now suppose that there exists a positive integer $t$ such that
\begin{equation}\label{f-14}
    \sum_{g\in \Phi_k}\lambda(g)=0 \mbox{ for all } k>t.
\end{equation}
Then $L(z)$ is a complex polynomial of degree $\leq t$ with constant $1$, so that we have
\begin{equation}\label{f-15}
  L(z)=(1-\omega_1z)(1-\omega_2z)\cdots(1-\omega_tz)
\end{equation}
with complex numbers $\omega_1,\omega_2,\cdots,\omega_t$. It follows that
\begin{equation}\label{f-16}
    L_s=-\omega_1^s-\omega_2^s-\cdots-\omega_t^s \mbox{ for all }s\geq 1.
\end{equation}


\section{A recursive formula and an estimate for a specific exponential sum}

Now we consider the exponential sums
\begin{equation}\label{f-17}
   G_u(a,b)=\sum_{c\in \mathbb{F}^*_q}\chi(ac^u+bc^{-1}), a\in \mathbb{F}^*_q,b\in \mathbb{F}_q,
\end{equation}
here $u\geq 1$ is a positive integer.

If $u=1$, then $G_u(a,b)$ is the well-known Kloosterman sum. If $u=3$, then $G_u(a,b)$ is called the {\it inverse cubic sum}. It is easily seen that $G_u(a,b)=G_u(ab^u,1)$ if $b\neq 0$.

In what follows, we proceed to give a recursive formula and an estimate of the exponential sum $G_u(a,b)$ by using the ideal introduced in subsection 2.3. To this end, we should define a multiplicative function $\lambda$ from the set $\Phi$ of all the monic polynomials in $\mathbb{F}_q[x]$ to the set of complex numbers.

We put $\lambda(1)=1$ and if $g\in \Phi_k$, $k\geq 1$, say
\begin{equation*}
    g(x)=\sum_{j=0}^k(-1)^jc_jx^{k-j} \mbox{ with }c_0=1,
\end{equation*}
suppose that $g$ factors as
\begin{equation*}
    g(x)=(x-\alpha_1)(x-\alpha_2)\cdots(x-\alpha_k)
\end{equation*}
in its splitting field, we set
\begin{equation*}
    \tilde{g}(x)=(x-\alpha_1^u)(x-\alpha_2^u)\cdots(x-\alpha_k^u)=\sum_{j=0}^k(-1)^j\tilde{c}_jx^{k-j} \mbox{ with }\tilde{c}_0=1.
\end{equation*}
It is easily seen that the numbers
\begin{equation*}
     \tilde{c}_1=\sum_{j=1}^k\alpha_j^u, c_{k-1}c_k^{-1}=\sum_{j=1}^k\alpha_j^{-1}
\end{equation*}
are invariant under the Frobenius automorphism $x\mapsto x^q$, thus $\tilde{c}_1, c_{k-1}c_k^{-1}\in \mathbb{F}_q$, and we can define
\begin{equation*}
    \lambda(g)=\chi(a\tilde{c}_1+bc_{k-1}c_k^{-1}) \mbox{ if } c_k\neq 0,
\end{equation*}
and $\lambda(g)=0$ if $c_k=0$.

Now we check that $\lambda(gh)=\lambda(g)\lambda(h)$ holds for all $g,h\in \Phi$. Suppose that
\begin{equation*}
     g(x)=(x-\alpha_1)(x-\alpha_2)\cdots(x-\alpha_k), h(x)=(x-\gamma_1)(x-\gamma_2)\cdots(x-\gamma_l),
\end{equation*}
and then
\begin{eqnarray*}
  && \tilde{g}(x)=(x-\alpha_1^u)(x-\alpha_2^u)\cdots(x-\alpha_k^u) = \sum_{j=0}^k(-1)^j\tilde{c}_jx^{k-j} \\
  &&\tilde{h}(x)=(x-\gamma_1^u)(x-\gamma_2^u)\cdots(x-\gamma_k^u) = \sum_{i=0}^l(-1)^i\tilde{d}_ix^{k-i}
\end{eqnarray*}
with
\begin{equation*}
   \tilde{ c}_1=\sum_{j=1}^k\alpha_j^u, \tilde{d}_1=\sum_{i=1}^l\gamma_i^u; c_{k-1}c_k^{-1}=\sum_{j=1}^k\alpha_j^{-1}, d_{l-1}d_l^{-1}=\sum_{i=1}^l\gamma_i^{-1}.
\end{equation*}
Therefore, if $$g(x)h(x)=\sum_{m=0}^{k+l}(-1)^mE_mx^{k+l-m}, \tilde{g}(x)\tilde{h}(x)=\sum_{m=0}^{k+l}(-1)^m\tilde{E}_mx^{k+l-m},$$
we have,
\begin{equation*}
    \tilde{E}_1=\tilde{c}_1+\tilde{d}_1, E_{k+l-1}E_{k+l}^{-1}=\sum_{j=1}^k\alpha_j^{-1}+\sum_{i=1}^l\gamma_i^{-1}=c_{k-1}c_k^{-1}+ d_{l-1}d_l^{-1}.
\end{equation*}
Thus we get that $\lambda(gh)=\lambda(g)\lambda(h)$ holds for all $g,h\in \Phi$.


By Newton's identities (\ref{f-7}), we have
\begin{equation*}
   \tilde{c}_1=\sum_{j=1}^k\alpha_j^u=p_u=\sum_{j=1}^{u-1}(-1)^je_jp_{u-j}+(-1)^{u-1}ue_u, u\geq 2,
\end{equation*}
where $p_j=\alpha_1^j+\cdots+\alpha_k^n$ is the power sum, since $p_j$ is a symmetric function on $\alpha_1,\cdots,\alpha_k$, we know that $p_j\in \mathbb{F}_q$, and $e_j=e_j(\alpha_1,\cdots,\alpha_k)=c_j$ for $j\leq k$ and $e_j=0$ for $j>k$. Therefore,
\begin{equation}\label{f-26}
   \tilde{c}_1=\left\{\begin{array}{ll}
                        \sum_{j=1}^k(-1)^jc_jp_{u-j} & \mbox{ if }k<u, \\
                        \sum_{j=1}^{u-1}(-1)^{j-1}c_jp_{u-j}+(-1)^{u-1}uc_u & \mbox{ if }k\geq u.
                      \end{array}
   \right.
\end{equation}
It is easily seen that $\sum_{j=1}^{u-1}(-1)^{j-1}c_jp_{u-j}=\pi(c_1,\cdots,c_{u-1})$ is a function on $c_1,\cdots,c_{u-1}$.

Hence, for $k>u+1$, if $\gcd(u,q)=1$, we can write
\begin{eqnarray*}
  \sum_{g\in \Phi_k}\lambda(g)&=& \sum_{c_1,c_2,\cdots,c_{k-1}\in \mathbb{F}_q}\sum_{c_k\in \mathbb{F}_q^*}\chi(a\tilde{c}_1+bc_{k-1}c_k^{-1})\\
  &=&q^{k-u-1}\left(\sum_{c_u\in \mathbb{F}_q}\chi(a(-1)^{u-1}uc_u)\right)\cdot\\
  & &\left(\sum_{c_k\in \mathbb{F}_q^*}\sum_{c_1, \cdots,c_{u-1}, c_{k-1}\in \mathbb{F}_q}\chi(a\sum_{j=1}^{u-1}(-1)^{j-1}c_jp_{u-j})\chi(bc_k^{-1}c_{k-1})\right)\\
  &=&0;
\end{eqnarray*}
If $b\neq 0$, we can write
\begin{eqnarray*}
  \sum_{g\in \Phi_k}\lambda(g)&=& \sum_{c_1,c_2,\cdots,c_{k-1}\in \mathbb{F}_q}\sum_{c_k\in \mathbb{F}_q^*}\chi(a\tilde{c}_1+bc_{k-1}c_k^{-1})\\
  &=&q^{k-u-1}\sum_{c_k\in \mathbb{F}_q^*}\sum_{c_1, \cdots,c_{u}\in \mathbb{F}_q}\chi\left(a\sum_{j=1}^{u-1}(-1)^{j-1}c_jp_{u-j}+(-1)^{u-1}uc_u\right)\\
  & &\left(\sum_{c_{k-1}\in \mathbb{F}_q}\chi(bc_k^{-1}c_{k-1})\right)\\
  &=&0;
\end{eqnarray*}
We note that the aim of presupposition $k>u+1$ is to separate $c_u$ and $c_{k-1}$ such that $c_u$ or $c_{k-1}$ is free in the above summation.

Therefore (\ref{f-14}) holds with $t=u+1$. From (\ref{f-15}) we obtain
\begin{equation*}
    L(z)=(1-\omega_1z)(1-\omega_2z)\cdots(1-\omega_{u+1}z)
\end{equation*}
for some complex numbers $\omega_1,\cdots,\omega_{u+1}$.

Now we calculate $L_s$ from (\ref{f-13}). Let $g$ be a monic irreducible polynomial in $\mathbb{F}_q[x]$ whose degree
divides $s$, and let $\gamma\in E=\mathbb{F}_{q^s}$ be a root of $g$. Then $g^{s/{\deg(g)}}$ is the characteristic polynomial of $\gamma$ over
$\mathbb{F}_q$; that is,
\begin{equation*}
    g(x)^{s/\deg(g)}=(x-\gamma)(x-\gamma^q)\cdots(x-\gamma^{q^{s-1}})=x^s-c_1x^{s-1}+\cdots+(-1)^{s-1}c_{s-1}x+(-1)^sc_s,
\end{equation*}
say. Then $c_1={\rm Tr}_{E/\mathbb{F}_q}(\gamma)$, $c_s=\gamma \gamma^q\cdots\gamma^{q^{s-1}}$, and $$c_{s-1}c_s^{-1}=\gamma^{-1}+\gamma^{-q}+\cdots+\gamma^{-q^{s-1}}={\rm Tr}_{E/\mathbb{F}_q}(\gamma^{-1}).$$
Thus we have
\begin{eqnarray*}
  \lambda\left(g(x)^{s/\deg(g)}\right)&=& \chi\left(a(\gamma^u+\gamma^{uq}+\cdots+\gamma^{uq^{s-1}})+b(c_{s-1}c_s^{-1}) \right)\\
  &=&\chi\left(a{\rm Tr}_{E/\mathbb{F}_q}(\gamma^u)+b{\rm Tr}_{E/\mathbb{F}_q}(\gamma^{-1})\right)\\
  &=&\chi^{(s)}(a\gamma^u+b\gamma^{-1}),
\end{eqnarray*}
and so
\begin{equation*}
    L_s=\sum_{g}\ ^*\deg(g)\lambda(g^{s/\deg(g)})=\sum_{g}\ ^*\sum_{\gamma\in E,g(\gamma)=0}\chi^{(s)}(a\gamma^u+b\gamma^{-1}),
\end{equation*}
where $^*$ means the summation is extend to all monic irreducible polynomial $g(x)\in \mathbb{F}_q[x]$
with $g(x)=x$ is excluded. If $g$ runs through the range of summation above, then $\gamma$ runs exactly through all elements of $E^*$. Thus
\begin{equation*}
    L_s=\sum_{\gamma\in E^*}\chi^{(s)}(a\gamma^u+b\gamma^{-1}).
\end{equation*}
Therefore, we have proved the following:

{\thm \label{thm-3} Let $\mathbb{F}_q$ be a finite field and $u$ be a positive integer. Let $s$ be any positive integer, and
$\chi^{(s)}$ be the lifting of the additive character $\chi$ of $\mathbb{F}_q$. For any two elements $a,b\in \mathbb{F}_q$ with $a\neq 0$, define
\begin{equation}\label{f-27}
   G_u^{(s)}(a,b)=\sum_{c\in \mathbb{F}^*_{q^s}}\chi^{(s)}(ac^u+bc^{-1}).
\end{equation}
Then, if either $\gcd(u,q)=1$ or $b\neq 0$, there exist complex numbers $\omega_1,\omega_2,\cdots, \omega_{u+1}$, only depending on $u,a,b$ and $\chi$, such that for any positive integer $s$ we have
\begin{equation}\label{f-2}
   G_u^{(s)}(a,b) =-\omega_1^s-\omega_2^s-\cdots\omega_{u+1}^s.
\end{equation}}

Moreover, for the complex numbers $\omega_1,\cdots,\omega_{u+1}$, we have the following:

{\thm \label{thm-5}Let $u$ be a positive integer with $\gcd(u,q)=\gcd(u+1,q)=1$ and $ab\neq 0$. Then the complex numbers $\omega_1,\cdots,\omega_{u+1}$ in Theorem \ref{thm-3} are all of absolute $\leq \sqrt{q}$.}

Thus we have the following:

{\cor Let $u$ be a positive integer with $\gcd(u,q)=\gcd(u+1,q)=1$ and $ab\neq 0$, and let the exponential sum $G_u^{(s)}(a,b)$ be defined as in Theorem \ref{thm-3}. Then we have
\begin{equation*}
    |G_u^{(s)}(a,b)|\leq (u+1)\sqrt{q} \mbox{ for all }s=1,2,\cdots.
\end{equation*}}

For the proof of Theorem \ref{thm-5}, we need the following lemma:

{\lem \label{lem-1} \cite[Lemma 6.55, p.310]{lidl} Let $\omega_1,\cdots,\omega_n$ be complex numbers, and let $B>0$, $C>0$ be constants such that
\begin{equation}\label{f-19}
   |\omega_1^s+\cdots+\omega_n^s|\leq CB^s \mbox{ for } s=1,2,\cdots.
\end{equation}
Then $|\omega_j|\leq B$ for $j=1,2,\cdots,n.$}


Now we give the proof of Theorem \ref{thm-5}.

\begin{proof}For any $a, b\in \mathbb{F}_q$ with $a\neq 0$, we write
\begin{equation*}
   \sum_{c\in E^*}\chi^{(s)}(ac^u+bc^{-1})=\sum_{c\in E^*}\chi\left({\rm Tr}_{E/\mathbb{F}_q}(ac^u+bc^{-1})\right)=\sum_{v\in \mathbb{F}_q}N(v)\chi(v),
\end{equation*}
where
\begin{equation*}
    N(v)=|\{c\in E^*|{\rm Tr}_{E/\mathbb{F}_q}(ac^u+bc^{-1})=v\}|.
\end{equation*}
If $\beta_0$ is a fixed element in $E$ with ${\rm Tr}_{E/\mathbb{F}_q}(\beta_0)=v$, then by Hilbert's Theorem 90, we know that
${\rm Tr}_{E/\mathbb{F}_q}(ac^u+bc^{-1})=v$ if and only if $ac^u+bc^{-1}=\beta^q-\beta+\beta_0$ for some $\beta\in E$, (see \cite[Theorem 2.25]{lidl}), or equivalently, $ac^{u+1}-(\beta^q-\beta+\beta_0)c+b=0$. Let $N$ be the number of solutions of
\begin{equation}\label{f-21}
  f(x,y)= ay^{u+1}-(x^q-x+\beta_0)y+b=0
\end{equation}
in $E^2$.

We proceed to show that $f(x,y)$ is absolutely irreducible over $\mathbb{F}_{q^s}$. We note that a polynomial is called {\it absolute irreducible} over a field $L$ means that this polynomial is irreducible over every extension field of $L$.

If there is an extension field $K$ of $\mathbb{F}_{q^s}$ such that $f(x,y)$ is reducible in $K[x,y]$. Let
  \begin{equation*}
    \Omega=\{g(x,y)| g(x,y) \mbox{ is an irreducible factor of } f(x,y) \mbox{ in } K[x,y]\}.
\end{equation*}
We define the action of the additive group of $\mathbb{F}_q$ on $\Omega$ as follows.
If $f_1(x,y)=a_0(y)+a_1(y)x+\cdots+a_t(y)x^t \in \Omega$, for every element $\xi\in \mathbb{F}_q$, define the action of $\xi$ on $f_1(x,y)$ by $\sigma_\xi: x\mapsto x+\xi,y\mapsto y$. It is clear that this is a group action, so either

Case (1) $f_1(x,y)$ is invariant under the action of every $\sigma_\xi, \xi\in \mathbb{F}_q$, or

Case (2) there are some orbits, such that the product of those polynomials in each orbit is invariant under the action of every $\sigma_\xi, \xi\in \mathbb{F}_q$.

Thus, both in case (1) and case (2), we always have a factor of $f(x,y)$, say $f_1(x,y)$, which is invariant under the action of $\mathbb{F}_q$.

We denote by $\partial_x(f_1(x,y))$ (resp. $\partial_y(f_1(x,y))$) the highest degree of $x$ (resp. $y$) among all monomials in $f_1(x,y)$, then all the polynomials in an orbit have the same $\partial_x$, and the same $\partial_y$.

We proceed to prove that $\partial_x(f_1(x,y))=q$.

Since $f_1(x,y)$ is invariant under the action of $\sigma_\xi,\xi\in \mathbb{F}_q$, considered as a polynomial in indeterminate $z$, the following polynomial
\begin{equation*}
    T(z)=f_1(x_0+z,y_0)-f_1(x_0,y_0)
\end{equation*}
has at least $q$ roots, namely, all the elements in $\mathbb{F}_q$, where $x_0,y_0$ are prescribed two elements arbitrarily. Thus $T(z)$ is divisible by $\prod_{\xi\in \mathbb{F}_q}(z-\xi)=z^q-z$. In other words,
\begin{equation*}
    T(z)=f_1(x_0+z,y_0)-f_1(x_0,y_0)=(z^q-z)q(z) \mbox{ for a } q(z)\in K[z].
\end{equation*}
Comparing the degrees of both sides yields that $\partial_z(q(z))=0$.
Hence $f_1(x,y)$ should have the form $g_1(y)+(x^q-x)g_2(y)$ for some polynomials $g_1(y),g_2(y)$. Suppose that $f(x,y)$ factors as $f(x,y)=f_1(x,y)f_2(y)$ with $\deg(f_2(y))\geq 1$.
Then we have
\begin{equation*}
    f(x,y)=ay^{u+1}-(x^q-x+\beta_0)y+b=(g_1(y)+(x^q-x)g_2(y))f_2(y),
\end{equation*}
which yields that $g_2(y)$ is a constant and $\deg(f_2(y))=1$. Suppose that $g_2(y)=\alpha_0, f_2(y)=\epsilon_0+\epsilon_1y$, then
\begin{equation*}
    ay^{u+1}-(x^q-x+\beta_0)y+b=g_1(y)(\epsilon_0+\epsilon_1y)+(x^q-x)\alpha_0(\epsilon_0+\epsilon_1y)
\end{equation*}
which implies that $\epsilon_0=0$ and thus $b=0$. Contrary to the fact that $b\neq 0$.

It follows that $f(x,y)$ can not have a proper factor which is invariant under the action of $\mathbb{F}_q$. Therefore, $f(x,y)$ can not factor as $f_1(x,y)g(y)$ for a $g(y)\in K[y]$ with $\deg(g(y))\geq 1$. By the same reason, $f(x,y)$ can not factor as $f_1(x,y)g(x)$ for a $g(x)\in K[x]$ with $\deg(g(x))\geq 1$, for if it does, it is obvious that $\partial_y(\sigma_\xi(f_1(x,y)))=\partial_y(f_1(x,y))$, then $f_1(x,y)$ is a proper factor of $f(x,y)$ which is invariant under the action of $\mathbb{F}_q$, a contradiction.

Therefore, we know that the group action $(\mathbb{F}_q,\Omega)$ has just one orbit, since the product of polynomials in an orbit is invariant under the action of $\mathbb{F}_q$. Thus we know that the group action $(\mathbb{F}_q,\Omega)$ is transitive and $|\Omega|$ divides $q$. Thus for every $f_1(x,y),f_2(x,y)\in \Omega$, we have $\partial_x(f_1(x,y))=\partial_x(f_2(x,y))=d_x$, $\partial_y(f_1(x,y))=\partial_y(f_1(x,y))=d_y$. So that $u+1=d_y |\Omega|$ and $q=d_x|\Omega|$. By the assumption that $\gcd(u+1,q)=1$, we get at last that $|\Omega|=1$ and $f(x,y)$ is irreducible in $K[x,y]$. This contradiction means that $f(x,y)$ is absolutely irreducible.

Now, by a famous result of Weil\footnote{It says that if a polynomial $f(x,y)\in \mathbb{F}_q[x]$ is absolutely irreducible, then there exists a constant $C$ such that the number $N$ of the solutions of the equation $f(x,y)=0$ satisfies $|N-q|\leq C\sqrt{q}$.}, see Weil \cite{weil1,weil2},
we have,
\begin{equation}\label{f-24}
    |N-q^s|\leq Cq^{s/2} \mbox{ for some constant }C,
\end{equation}
where the constant $C$ is independent on $s$.

On the other hand, for each fixed $y\in E$ satisfies (\ref{f-21}), there are $q$ choices of $x$, namely by adding any element of $\mathbb{F}_q$, thus we have
\begin{equation}\label{f-23}
  N=qN(v)
\end{equation}



From (\ref{f-24}),(\ref{f-23}), it follows that
\begin{equation*}
  | N(v)-q^{s-1}|\leq Cq^{s/2-1}.
\end{equation*}
Denote by $R(v)$ the number $N(v)-q^{s-1}$, then $|R(v)|\leq Cq^{s/2-1}$, hence
\begin{eqnarray*}
  |G_u^{(s)}(a,b)|&=&|-\omega_1^s-\cdots-\omega_{u+1}^s| \\
 &=& \left|\sum_{c\in E^*}\chi^{(s)}(ac^u+bc^{-1})\right|=\left|\sum_{v\in \mathbb{F}_q}N(v)\chi(v)\right|\\
 &=&\left|\sum_{v\in \mathbb{F}_q}(q^{s-1}+R(v))\chi(v)\right|=\left|\sum_{v\in \mathbb{F}_q}R(v)\chi(v)\right|\\
 &\leq& Cq^{s/2}.
\end{eqnarray*}
The desired result then follows with Lemma \ref{lem-1}. This completes the proof.
\end{proof}

Denote $G_u^{(s)}(a,b)$ by $G^{(s)}$, then by Newton's identities, or by Dickson polynomials, we have the following:

{\cor \label{cor-3.5}Let $G^{(s)}$ and the numbers $\omega_1,\cdots,\omega_{u+1}$ be defined as above. Then
\begin{equation*}
    G^{(s)}=\sum_{j=1}^{s-1}(-1)^{j-1}e_j(\omega_1,\cdots,\omega_{u+1})G^{(s-j)}+(-1)^se_s(\omega_1,\cdots,\omega_{u+1}) \mbox{ for all }s\geq 2.
\end{equation*}

}
Because $e_s(\omega_1,\cdots,\omega_{u+1})=0$ whenever $s>u+1$. In order to setting the recursive formulae to work, we only need to know the $u+1$ initial values $G^{(1)},\cdots,G^{(u+1)}$, since we can find the elementary functions values $e_1,\cdots,e_{u+1}$ by (\ref{f-9}) under the presupposition that the characteristic of the finite field is bigger than $u+1$. On the other hand, if we can find the values of $\sum_{g\in \Phi_1}\lambda(g), \cdots,\sum_{g\in \Phi_{u+1}}\lambda(g)$, we can also determine the recursive formula without any restriction on the characteristic of the finite field. For example, when $u=1$, by Theorem \ref{thm-1}, we obtain the following well-known results about the Kloosterman sums. See \cite[p.226]{lidl} for instance.

{\cor Let $a,b$ be two elements in a finite field $\mathbb{F}_q$. Then there are two complex numbers $\omega_1,\omega_2$, depending on $ab$ and $\chi$,
 such that for every positive integer $s$, we have
 \begin{equation}\label{f-25}
    k(\chi^{(s)},a,b)=\sum_{c\in \mathbb{F}^*_{q^s}}\chi^{(s)}(ac+bc^{-1})=-\omega_1^s-\omega_2^s.
 \end{equation}
 Moreover, we have $|\omega_1|=|\omega_2|=\sqrt{q}$.}

 Since $\omega_1+\omega_2=-k(\chi,a,b)$, $ \omega_1\omega_2=q$, we have $e_1(\omega_1,\omega_2)=-k(\chi,a,b),e_2(\omega_1,\omega_2)=q$, by
 Newton's identities or by property of Dickson polynomials, we get the following:
 {\cor \cite[p.229]{lidl} Denote $k(\chi^{(s)},a,b)$ by $k^{(s)}$. Then
 \begin{equation}\label{f-26}
    k^{(s)}=-k^{(s-1)}k-k^{(s-2)} \mbox{ for }s\geq 2,
 \end{equation}
 and
 \begin{equation}\label{f-27}
   k^{(s)}=-\left(\omega_1^s+\left(\frac{q}{\omega_1}\right)^s\right)=-D_s^{(1)}(\omega_1+\frac{q}{\omega_1},q)=-D_s^{(1)}(-k,q).
 \end{equation} where we put $k^{(0)}=-2, k^{(1)}=k=k(\chi,a,b)$, $D_s^{(1)}(.,.)$ is the Dickson polynomial.}
\vskip 0.3 cm
 If $u=2$ and $q$ is even, we proceed to compute the corresponding function $L(z)$ as follows.

 Obviously, if $ab\neq 0$, then
 \begin{equation*}
    \sum_{g\in \Phi_1}\lambda(g)=\sum_{g=x-c,c\neq 0}\lambda(g)=\sum_{c\neq 0}\chi(ac^2+bc^{-1})=G_2(a,b).
 \end{equation*}
 Moreover,
 \begin{eqnarray*}
 \sum_{g\in \Phi_2}\lambda(g)&=&\sum_{c_1\in \mathbb{F}_q,c_2\in \mathbb{F}^*_q}\chi(a(\alpha_1^2+\alpha_2^2)+bc_1c_2^{-1})\\
  &=&\sum_{c_1\in \mathbb{F}_q,c_2\in \mathbb{F}^*_q}\chi(a(\alpha_1+\alpha_2)^2+bc_1c_2^{-1})\\
  &=&\sum_{c_1\in \mathbb{F}_q,c_2\in \mathbb{F}^*_q}\chi(ac_1^2+bc_1c_2^{-1}),
 \end{eqnarray*}
where $\alpha_1,\alpha_2$ are the roots of $x^2-c_1x+c_2=0$ in an extension of the finite field $\mathbb{F}_q$. Thus
\begin{equation*}
    \sum_{g\in \Phi_2}\lambda(g)=q+\sum_{c_1\in \mathbb{F}^*_q}\chi(ac_1^2)\sum_{c_2\in \mathbb{F}^*_q}\chi(bc_1c_2^{-1})=q+1.
\end{equation*}
Furthermore,
\begin{eqnarray*}
   \sum_{g\in \Phi_3}\lambda(g)&=& \sum_{g(x)=x^3-c_1x^2+c_2x-c_3, c_3\neq 0}\lambda(g) \\
  &=& \sum_{c_1,c_2\in \mathbb{F}_q}\sum_{c_3\in \mathbb{F}^*_q}\chi(a(\alpha_1^3+\alpha_2^3+\alpha_3^3)+bc_2c_3^{-1})\\
  &=&\sum_{c_1,c_2\in \mathbb{F}_q}\sum_{c_3\in \mathbb{F}^*_q}\chi(a(c_1^3-c_1c_2+c_3)+bc_2c_3^{-1})\\
  &=&\sum_{c_1\in \mathbb{F}_q,c_3\in \mathbb{F}^*_q}\chi(a(c_1^3+c_3)\sum_{c_2\in \mathbb{F}_q}\chi(c_2(bc_3^{-1}-ac_1))\\
  &=&q\sum_{c_1,c_3\in \mathbb{F}^*_q, bc_3^{-1}=ac_1}\chi(ac_1^3+c_3)\\
  &=&qG_3(a,a^{-1}b).
\end{eqnarray*}
Thus, we have
\begin{equation*}
    L(z)=1+G_2(a,b)z+(q+1)z^2+qG_3(a,a^{-1}b)z^3.
\end{equation*}
By Corollary \ref{cor-3.5}, we have the following:
{\prop Let $q$ be a power of $2$, $\mathbb{F}_q$ be a finite field.  For any integer $s$, define
\begin{equation*}
    G^{(s)}(a,b)=\sum_{c\in \mathbb{F}^*_{q^s}}\chi^{(s)}(ac^2+bc^{-1}).
\end{equation*}
Then,

(1) for every elements $a,b\in \mathbb{F}^*_q$, we have
\begin{equation*}
  G^{(s)}(a,b)=-G_2(a,b)G^{(s-1)}(a,b)-(q+1)G^{(s-2)}-qG_3(a,a^{-1}b)G^{(s-3)} \mbox{ for all }s>3,
\end{equation*}
where $G_u(a,b)$ is defined by (\ref{f-17}).

(2) If $1+G_2(a,b)z+(q+1)z^2+qG_3(a,a^{-1}b)z^3=(1-\omega_1z)(1-\omega_2z)(1-\omega_3z)$, then for every positive integer $s$,
\begin{equation*}
    G^{(s)}(a,b)=-\omega_1^s-\omega_2^s-\omega_3^s.
\end{equation*}
}
\vskip 0.1 cm
In the case of $u=2$ and $q$ is odd, the situation is slightly different, we should distinguish the cases of ${\rm char}(\mathbb{F}_q)=3$ and ${\rm char}(\mathbb{F}_q)>3$. In these two cases, we have
 \begin{equation*}
    \sum_{g\in \Phi_1}\lambda(g)=\sum_{g=x-c,c\neq 0}\lambda(g)=\sum_{c\neq 0}\chi(ac^2+bc^{-1})=G_2(a,b),
 \end{equation*}
and
\begin{eqnarray*}
  \sum_{g\in \Phi_2}\lambda(g)&=&\sum_{c_1,c_2\in \mathbb{F}_q,c_2\neq 0}\chi(a(\alpha_1^2+\alpha_2^2)+bc_1c_2^{-1}) \\
 &=&\sum_{c_2\neq 0}\chi(-2ac_2)\sum_{c_1\in \mathbb{F}_q}\chi(ac_1^2+bc_1c_2^{-1}).
\end{eqnarray*}
Since $ac_1^2+bc_1c_2^{-1}$ is a quadratic function in $c_1$, the inner sum is
\begin{equation*}
    \sum_{c_1\in \mathbb{F}_q}\chi(ac_1^2+bc_1c_2^{-1})=\chi(-4^{-1}b^2c_2^{-2}a^{-1})\eta(a)g(\eta,\chi),
\end{equation*}
where $\eta$ is the quadratic charter, and $g(\eta,\chi)$ is the Gaussian sum
of the quadratic character $\eta$ and the additive character $\chi$. This Gaussian sum has been determined explicitly, see \cite[Theorem 5.12, Theorem 5.15]{lidl} for instance.
Hence we have
\begin{eqnarray*}
   \sum_{g\in \Phi_2}\lambda(g)&=& \eta(a)g(\eta,\chi)\sum_{c_2\neq 0}\chi(-2ac_2)\chi(-4^{-1}b^2c_2^{-2}a^{-1}) \\
 &=& \eta(a)g(\eta,\chi)\sum_{c_2\neq 0}\chi(-4^{-1}a^{-1}b^2c_2^2-2ac_2^{-1})\\
 &=&\eta(a)g(\eta,\chi)G_2(-\frac{b^2}{4a},-2a).
\end{eqnarray*}
If ${\rm char}(\mathbb{F}_q)=3$, then
\begin{eqnarray*}
 \sum_{g\in \Phi_3}\lambda(g)&=&\sum_{c_1,c_2,c_3\in \mathbb{F}_q,c_3\neq 0}\chi(a(\alpha_1^3+\alpha_2^3+\alpha_3^3)+bc_2c_3^{-1}) \\
 &=&\sum_{c_1,c_2,c_3\in \mathbb{F}_q,c_3\neq 0}\chi(ac_1^3+bc_2c_3^{-1})=0.
\end{eqnarray*}
If ${\rm char}(\mathbb{F}_q)>3$, then
\begin{eqnarray*}
 \sum_{g\in \Phi_3}\lambda(g)&=&\sum_{c_1,c_2,c_3\in \mathbb{F}_q,c_3\neq 0}\chi(a(\alpha_1^3+\alpha_2^3+\alpha_3^3)+bc_2c_3^{-1}) \\
 &=&\sum_{c_1,c_2,c_3\in \mathbb{F}_q,c_3\neq 0}\chi(a(c_1^3-3c_1c_2+3c_3)+bc_2c_3^{-1})\\
 &=&\sum_{c_1,c_3\in \mathbb{F}_q,c_3\neq 0}\chi(ac_1^3+3c_3)\sum_{c_2\in \mathbb{F}_q}\chi((bc_3^{-1}-3ac_1)c_2)\\
 &=&q\sum_{c_1\in \mathbb{F}_q^*}(ac_1^3+a^{-1}bc_1^{-1})\\
 &=&qG_3(a,a^{-1}b).
\end{eqnarray*}
Therefore, we obtain the following two results.

{\prop Let $q$  be a power of $3$.  For any integer, define
\begin{equation*}
    G^{(s)}(a,b)=\sum_{c\in \mathbb{F}^*_{q^s}}\chi(ac^2+bc^{-1}).
\end{equation*}
Then,

(1) for every elements $a,b\in \mathbb{F}^*_q$, we have
\begin{equation*}
  G^{(s)}(a,b)=-G_2(a,b)G^{(s-1)}(a,b)-\eta(a)g(\eta,\chi)G_2(-\frac{b^2}{a},a)G^{(s-2)}(a,b)
\end{equation*}
 for all $s>2$, where $G_u(a,b)$ is defined by (\ref{f-17}).

(2) If $1+G_2(a,b)z+\eta(a)g(\eta,\chi)G_2(-\frac{b^2}{a},a)z^2=(1-\omega_1z)(1-\omega_2z)$, then for every positive integer $s$,
\begin{equation*}
    G^{(s)}(a,b)=-\omega_1^s-\omega_2^s=-D_s^{(1)}\left(-G_2(a,b),\eta(a)g(\eta,\chi)G_2(-\frac{b^2}{a},a)\right).
\end{equation*}
}

{\prop Let $\mathbb{F}_q$ be a finite field with ${\rm char}(\mathbb{F}_q)>3$.  For any integer, define
\begin{equation*}
    G^{(s)}(a,b)=\sum_{c\in \mathbb{F}^*_{q^s}}\chi^{(s)}(ac^2+bc^{-1}).
\end{equation*}
Then,

(1) for every elements $a,b\in \mathbb{F}^*_q$, we have
\begin{eqnarray*}
  G^{(s)}(a,b)&=&-G_2(a,b)G^{(s-1)}(a,b)-\eta(a)g(\eta,\chi)G_2(-\frac{b^2}{4a},-2a)G^{(s-2)}(a,b)\\
  & &-qG_3(a,a^{-1}b)G^{(s-3)}(a,b)
\end{eqnarray*}
 for all $s>3$.

(2) If $1+G_2(a,b)z+\eta(a)g(\eta,\chi)G_2(-\frac{b^2}{4a},-2a)z^2+qG_3(a,a^{-1}b)z^3=(1-\omega_1z)(1-\omega_2z)(1-\omega_3z)$, then for every positive integer $s$,
\begin{equation*}
    G^{(s)}(a,b)=-\omega_1^s-\omega_2^s-\omega_3^s.
\end{equation*}
Where $G_u(a,b)$ is defined by (\ref{f-17}).
}

\section{Sequences constructed from the exponential sums}
In this section, we restrict the characteristic of the finite field being $2$, and we denote $G_u(a,1)$ simply by $G_u(a)$. For every element $a\in \mathbb{F}^*_q$, we define a sequence $\mathcal{G}_a$ by
\begin{equation*}
    \mathcal{G}_a=\left(G_u(ax)\right)_{x\in \mathbb{F}^*_q},
\end{equation*}
The correlation of two such sequences is defined
as
\begin{equation}\label{f-41}
   \mathcal{C}_{a,b}=
   C_{\mathcal{G}_a,\mathcal{G}_b}=\sum_{x\in
   \mathbb{F}^*_{q}}G_u(ax)G_u(bx).
\end{equation}
The autocorrelation of a sequence $\mathcal{G}_a$ is defined as
\begin{equation}\label{f-42}
   \mathcal{A}_{\mathcal{G}_a}(h)= \sum_{x\in
   \mathbb{F}^*_{q}}G_u(ax)G_u(ahx), h\in \mathbb{F}^*_{q}.
\end{equation}

About the distribution of values of the autocorrelation of the sequence $\mathcal{G}_a$, we have the following:

{\prop For every $a\in \mathbb{F}^*_q$ and a positive integer $u$ coprime with $(q-1)$, we have
\begin{equation*}
    \mathcal{A}_{\mathcal{G}_a}(h)=\left\{\begin{array}{ll}
                                            -q-1, & \mbox{ if $h\neq 1$} \\
                                            q^2-q-1, & \mbox{ if $h=1$.}
                                          \end{array}
    \right.
\end{equation*}}
Therefore, $\mathcal{G}_a$ is a two-valued correlation sequence which has some applications in communications, for example, Code Division Multiple Access (CDMA) systems, etc.
\begin{proof}Direct evaluation shows that
\begin{eqnarray*}
   \mathcal{A}_{\mathcal{G}_a}(h)&=&\sum_{x\in \mathbb{F}^*_q}G_u(ax)G_u(ahx)\\
 &=& \sum_{x\in \mathbb{F}^*_q}\sum_{c\in \mathbb{F}^*_q}\chi(axc^u+c^{-1})\sum_{d\in \mathbb{F}^*_q}\chi(ahxd^u+d^{-1})\\
 &=&\sum_{c,d\in \mathbb{F}^*_q}\chi(c^{-1}+d^{-1}) \sum_{x\in \mathbb{F}^*_q}\chi(ax(c^u+hd^u))\\
 &=&\sum_{c,d\in \mathbb{F}^*_q}\chi(c^{-1}+d^{-1})\left( \sum_{x\in \mathbb{F}_q}\chi(ax(c^u+hd^u))-1\right)\\
 &=&q\sum_{d\in \mathbb{F}^*_q}\chi((h^{-u'}+1)d^{-1})-\sum_{c,d\in \mathbb{F}_q^*}\chi(c^{-1}+d^{-1}) \quad (uu'\equiv 1\ {\rm mod}\ q-1)\\
 &=&\left\{\begin{array}{ll}
                                            -q-1, & \mbox{ if $h\neq 1$} \\
                                            q^2-q-1, & \mbox{ if $h=1$,}
                                          \end{array}
    \right.
\end{eqnarray*}
which is the desired result.\end{proof}
Hence we obtain the following corollary:
{\cor\label{cor-43} For every pair of $(a,b)\in (\mathbb{F}^*_{q})^2$, and a positive integer $u$ coprime with $(q-1)$, we have
\begin{equation}\label{f-44}
   \sum_{x\in \mathbb{F}^*_{q}}G_u(ax)G_u(bx)=\left\{\begin{array}{ll}
              q^2-q-1,& \mbox{ if } a=b \\
              -q-1, & \mbox{ otherwise. }
            \end{array}
  \right.
\end{equation}}
We also
have the following proposition.
{\prop \label{p-3}
For every $a,b,c\in \mathbb{F}^*_{q}$, and a positive integer $u$ coprime with $(q-1)$, we
have
\begin{equation}\label{f-46}
   \sum_{x\in
   \mathbb{F}^*_{q}}G_u(ax)G_u(b(c-x))=q G_u(c(a^{u'}+b^{u'})^u)+G_u(bc),
\end{equation}
where $G_u(a)=G_u(a,1)$ is defined by (\ref{f-17}) and $uu'\equiv 1{\rm mod }\ q-1$.}
\begin{proof} By definition, we have
\begin{eqnarray*}
  &&\sum_{x\in
   \mathbb{F}^*_{q}}G_u(ax)G_u(b(c-x))
    = \sum_{x\in
   \mathbb{F}^*_q}\sum_{y,z\in
   \mathbb{F}^*_{q}}\chi\left(axy^u+y^{-1}+b(c-x)z^u+z^{-1}\right)\\
   &=&  \sum_{y,z\in
   \mathbb{F}^*_{q}}\chi(y^{-1}+z^{-1}+bcz^u)\left(\sum_{x\in \mathbb{F}_q}\chi\left(x(ay^u-bz^u)\right)-1\right)\\
   &=&q \sum_{z\in
   \mathbb{F}^*_{q}}\chi\left(bcz^u+(a^{u'}{b^{-u'}}+1)z^{-1}\right)-\sum_{z\in \mathbb{F}_q^*}\chi(bcz^u+z^{-1})\sum_{y\in \mathbb{F}_q^*}\chi(y^{-1})\\
   &=&q G_u(c(a^{u'}+b^{u'})^u)+G_u(bc).
\end{eqnarray*}This completes the proof.\end{proof}

\section*{Concluding remarks}

Firstly, for the exponential sum
   \begin{equation*}
    G_u(a,b)=\sum_{c\in \mathbb{F}^*_{q}}\chi(ac^u+bc^{-1}),
\end{equation*}
it is easily seen that
\begin{equation*}
    G_u(a,b)=\sum_{c\in \mathbb{F}^*_{q}}\chi(ac^{q-1-u}+bc).
\end{equation*}
However, in the extension field of $\mathbb{F}_q$, we should have
\begin{equation*}
  \sum_{c\in \mathbb{F}^*_{q^s}}\chi(ac^{q-1-u}+bc)\neq \sum_{c\in \mathbb{F}^*_{q^s}}\chi(ac^u+bc^{-1}).
\end{equation*}
Thus, by Theorem \ref{thm-1}, one can only obtain that there are $q-2-u$ complex numbers $\omega_1,\cdots,\omega_{q-2-u}$ such that $|\omega_j|=\sqrt{q}, j=1,\cdots,q-2-u$, and
\begin{equation*}
    \sum_{c\in \mathbb{F}_{q^s}}\chi^{(s)}(ac^{q-1-u}+bc)=-\omega_1^s-\cdots-\omega_{q-2-u}^s
\end{equation*}
for all $s=1,2,\cdots.$
This is obviously different with Theorem \ref{thm-3}. In other words, Theorem \ref{thm-3} is not covered by Theorem \ref{thm-1}.

There is a question still open now:

{\bf Open Question}: Does each of those complex numbers in Theorem \ref{thm-3} has magnitude $\sqrt{q}$ ?

About this question, we tend to have a positive answer. We think a possible approach to verifying it is using the theory of algebraic geometry.

Secondly, after we finished this paper, we found that one can generalize Theorem \ref{thm-3} to a more generic case as follows:

{\thm \label{thm-6}Let $\mathbb{F}_q$ be a finite field and $f(x),g(x)\in \mathbb{F}_q[x]$ be polynomials with degree $m\geq 1$ and $n\geq 1$, respectively. Let $s$ be any positive integer, and
$\chi^{(s)}$ be the lifting of the additive character $\chi$ of $\mathbb{F}_q$. Define
\begin{equation*}
   G^{(s)}(f,g)=\sum_{c\in \mathbb{F}_{q^s}}\chi^{(s)}(f(c)+g(c^{-1})).
\end{equation*}
Suppose that either $\gcd(m,q)=1$ or $\gcd(n,q)=1$. Then there exist complex numbers $\omega_1,\omega_2,\cdots, \omega_{m+n}$, only depending on $f,g$ and $\chi$, such that for any positive integer $s$ we have
\begin{equation*}
   G^{(s)}(f,g) =-\omega_1^s-\omega_2^s-\cdots-\omega_{m+n}^s.
\end{equation*}}
{\thm Let $f(x),g(x)\in \mathbb{F}_q[x]$ be two polynomials with degree $m\geq 1$ and $n\geq 1$, respectively. If either $\gcd(m,q)=1$ or $\gcd(n,q)=1$,
then the complex numbers $\omega_1,\cdots,\omega_{m+n}$ in Theorem \ref{thm-6} are all of absolute $\leq \sqrt{q}$ provided that $\gcd(m+n,q)=1$.}

We will report these results in a forthcoming paper.
\bibliographystyle{amsplain}

\begin{thebibliography}
\bibliography
\bibitem{coulter}Robert S. Coulter, Explicit evaluation of some Weil sums, Acta Arithmetica, {\bf 83}, 241-251(1998).
\bibitem{lidl}R. Lidl, H. Niederriter, Finite Fields, Encyclopedia Math.
Appl. Vol. {\bf 20}, Addison-Wesley, Reading, 1983.
\bibitem{LMT}R. Lidl, G.L. Mullen and G. Turnwald, Dickson
polynomials, Pitman Monographs and Surveys in Pure and Applied
Mathematics, {\bf 65}, Longman Group UK Limited 1993.


\bibitem{lison}Petr Lison\^{e}k, On the connection between Kloosterman sums and elliptic curves, S.W. Golomb et al. (Eds.): SETA 2008, LNCS 5203, pp. 182-187(2008).

\bibitem{moiso}Marko Moisio, Kalle Ranto, Kloosterman sum identities and low-weight codewords in a cyclic code with two zeros, Finite Fields and Their Appl. {\bf 13}, 922-935(2007).
\bibitem{vlugt}Marcel Van Der Vlugt, Hasse-Davenport curves, Gauss sums, and weight distributions of irreducible cyclic codes, Journal of Number Theory, {\bf 55}, 145-159(1996).

\bibitem{wan}D. Wan, Minimum polynomials and Distincness of Kloosterman sums, Finite Fields and Their Appl. {\bf 1}, 189-203(1995).
\bibitem{weil1} A. Weil, On the Reimann hypothsis in function fields, Proc. Nat. Acad. Sci. USA. {\bf 27}, 345-347(1941).
\bibitem{weil2}A. Weil, Sur les courbes alg\'{e}briques et les vari\'{e}t\'{e}s qui s'en d\'{e}duisent, Actualit\'{e}s Sci. Ind., no. {\bf 1041}, Hermann, Paris, 1948.
\end{thebibliography}

\end{document}